\documentclass{birkjour}
\usepackage[utf8]{inputenc}
\usepackage{amssymb,bbold}
\usepackage{tikz}

\newcommand{\ben}{\begin{enumerate}}  
\newcommand{\een}{\end{enumerate}}  

\newtheorem{theorem}{Theorem}
\newtheorem{proposition}[theorem]{Proposition}
\newtheorem{lemma}[theorem]{Lemma}

\newtheorem{conjecture}[theorem]{Conjecture}
\newtheorem{remark}[theorem]{Remark}
\newtheorem{example}[theorem]{Example}
\newcommand{\beq}{\begin{equation}}
\newcommand{\eeq}{\end{equation}}

\newcommand{\bex}{\begin{example}}
\newcommand{\eex}{\end{example}}

\def\bel{\begin{lemma}}
\def\eel{\end{lemma}}
\def\bet{\begin{theoreme}}
\def\eet{\end{theoreme}}
\def\bed{\begin{definition}}
\def\eed{\end{definition}}
\def\ber{\begin{remark}}
\def\eer{\end{remark}}

\def\rr{{\mathbb R}}
\def\zz{{\mathbb Z}}
\def\cc{{\mathbb C}}
\def\nn{{\mathbb N}}

\def\Im{\operatorname{Im}}
\def\Re{\operatorname{Re}}
\def\i{\mathrm{i}}
\def\Dom{\operatorname{Dom}}

\def\e{\mathrm{e}}

\def\r{{\rm r}}

\def\I{{\rm I}}

\def \p{\mathrm{p}}

\renewcommand{\c}{\mathrm{c}}

\renewcommand{\sp}{\operatorname{spec}}

\def\Ran{\operatorname{Ran}}

\def\one{\mathbb{1}}

\def\cD{{\mathcal D}}
\def\cF{{\mathcal F}}
 \def\cH{{\mathcal H}}
\def\cI{{\mathcal I}}
 \def\cK{{\mathcal K}}

\def\R{\rr}
\def\C{\cc}

\begin{document}

\title{Homogeneous rank one perturbations \\
  and inverse square potentials}
\author{Jan Dereziński}
\thanks{The financial support of the National Science
Center, Poland, under the grant UMO-2014/15/B/ST1/00126, is gratefully
acknowledged. The author is grateful to Serge~Richard, Vladimir~Georgescu and Laurent~Bruneau for discussions and collaboration.}
\address{Department of Mathematical Methods in Physics\\
Faculty of Physics\\
University of Warsaw\\
Pasteura 5\\
02-093 Warszawa, Poland}
\email{jan.derezinski@fuw.edu.pl}

\begin{abstract}
Following~\cite{D,BDG,DR},  I describe several exactly solvable families of closed operators on $L^2[0,\infty[$. Some of these families are defined by the theory of singular rank one perturbations.
The remaining families are  Schrödinger operators with inverse square potentials and various boundary conditions.
I describe a close  relationship between these families. In all of them one can observe interesting ``renormalization group flows'' (action of the group of dilations).
\end{abstract}
\maketitle

\section{Introduction}

My  contribution  consists of an introduction and 3 sections, each describing  an interesting family of  exactly solvable closed operators on $L^2[0,\infty[$.

The first two sections seem at first unrelated. Only in the third section the reader will see a relationship.

Section~\ref{sec1} is based on~\cite{D}. It is devoted to 
two families of operators, $H_{m,\lambda}$ and $H_0^\rho$, obtained by a rank one perturbation of a certain generic self-adjoint operator. The operators 
can be viewed as an elementary toy model illustrating some properties of the renormalization group. Note that in this section we do not use special functions. However we use a relatively sophisticated  
technique to define an operator, called sometimes {\em singular perturbation theory} or the {\em Aronszajn--Donoghue method}, see e.g.,~\cite{AK,KS,DF}.

Section~\ref{sec2} is based on my joint work with Bruneau and Georgescu~\cite{BDG}, and also with Richard~\cite{DR}. It is devoted to Schrödinger operators with potentials proportional to $\frac{1}{x^2}$. Both $-\partial_x^2$ and $\frac{1}{x^2}$ are homogeneous of degree $-2$. With appropriate homogeneous boundary conditions, we  obtain a family of operators $H_m$, which we call {\em homogeneous Schrödinger operators}. They are  also homogeneous of degree -2. One can compute all basic quantities for these operators using special functions--more precisely, {\em Bessel-type functions} and the {\em Gamma function}.

The operators $H_m$ are defined only for $\Re{m}>-1$. We conjecture that they cannot be extended to the left of the line $\Re m=-1$ in the sense described in our paper. This conjecture was stated in~\cite{BDG}.
It has not been proven or disproved so far.

Finally, Section~\ref{sec3} is based on my joint work with Richard~\cite{DR},  and also on~\cite{D}. It describes more general Schrödinger operators with the inverse square potentials. They are obtained
by  mixing the boundary conditions. These operators in general are no longer homogeneous, because  their homogeneity is (weakly) broken by their boundary condition---hence the name {\em almost homogeneous Schrödinger operators}. They can be organized in two families $H_{m,\kappa}$ and $H_0^\nu$.

It turns out that there exists a close relationship between the operators from Section~\ref{sec3} and from Section~\ref{sec1}: they are similar to one another. In particular, they have the same point spectrum.

Almost homogeneous Schrödinger operators in the self-adjoint case have been described in the literature before, see e.g.,~\cite{GTV}. However, the non-self-adjoint case seems to heve been first described in~\cite{DR}. A number of new exact formulas about these operators is contained in~\cite{BDG,PR,DR} and~\cite{D}.

Let us also mention one amusing observation, which seems to be original, about self-adjoint extensions of \[-\partial_x^2+\Big(-\frac14+\alpha\Big)\frac{1}{x^2}.\] The ``renormalization group'' acts on the set of these extensions, as described in a table
after Prop.~\ref{table0}. Depending on $\alpha\in\rr$, we obtain 4 ``phases'' of the problem. Some analogies to the condensed matter physics are suggested.

\section{Toy model of renormalization group}
\label{sec1}

Consider the Hilbert space $\cH= L^2[0,\infty[$ and the operator $X$
\[ Xf(x):=xf(x).\]

Let $m\in\cc$, $\lambda\in\cc\cup\{\infty\}$.
Following~\cite{D}, we  consider a family of operators formally given by
\beq H_{m,\lambda}:=X+\lambda |x^{\frac{m}{2}}\rangle\langle
x^{\frac{m}{2}}|.\label{pertu}\eeq

In the  perturbation $ |x^{\frac{m}{2}}\rangle\langle
x^{\frac{m}{2}}|$ we use the Dirac ket-bra notation, hopefully self-explanatory.  Unfortunately, the function $x\mapsto x^{\frac{m}{2}}$
is never square integrable. Therefore, this perturbation is never an operator. It can be however understood as a quadratic form. We will see below how to interpret
(\ref{pertu}) as an operator.

If $-1<\Re m<0$,
the perturbation 
$ |x^{\frac{m}{2}}\rangle\langle
x^{\frac{m}{2}}|$ is form bounded relatively to $X$,
and then  $H_{m,\lambda}$ can be defined by the form boundedness technique. The perturbation is formally rank one. Therefore, 
\begin{align*} &(z-H_{m,\lambda})^{-1}=(z-X)^{-1}\\
&+\sum_{n=0}^\infty
(z-X)^{-1}|x^{\frac{m}{2}}\rangle
(-\lambda)^{n+1}\langle x^{\frac{m}{2}}|(z-X)^{-1}|x^{\frac{m}{2}}\rangle^n
\langle x^{\frac{m}{2}}
|(z-X)^{-1}\\
&=(z-X)^{-1}\\
&+\Big(\lambda^{-1}-\langle x^{\frac{m}{2}}|(z-X)^{-1}|x^{\frac{m}{2}}\rangle\Big)^{-1}
(z-X)^{-1}|x^{\frac{m}{2}}\rangle
\langle x^{\frac{m}{2}}
|(z-X)^{-1}.
\end{align*}

It is an easy exercise in  complex analysis to compute
\begin{equation*}
\langle x^{\frac{m}{2}}|(z-X)^{-1}|x^{\frac{m}{2}}\rangle
=\int_0^\infty x^m(z-x)^{-1}\d x=
(-z)^{m}\frac{\pi}{\sin\pi
    m}.\end{equation*}
Therefore, the resolvent of
 $H_{m,\lambda}$  can be given in a closed form:
\begin{multline*} (z-H_{m,\lambda})^{-1}=(z-X)^{-1}\\
\hspace{-16ex}+\Big(\lambda^{-1}-(-z)^{m}\frac{\pi}{\sin\pi
    m}\Big)^{-1} (z-X)^{-1}|x^{\frac{m}{2}}\rangle\langle x^{\frac{m}{2}}
|(z-X)^{-1}.
\end{multline*}
The rhs of the above formula defines a function with values in bounded  operators satisfying the resolvant equation for all
$-1<\Re m<1$ and
 $\lambda\in\cc\cup\{\infty\}$. Therefore, the method of pseudoresolvent~\cite{Kato2} allows  us to define a holomorphic family of 
closed operators $H_{m,\lambda}$.
Note that
 $H_{m,0}=X$.

The case $m=0$ is special:  $H_{0,\lambda}=X$ for all $\lambda$.
One can however introduce another holomorphic family of operators $H_0^\rho$ for any $\rho\in\cc\cup\{\infty\}$ by
  \begin{equation*}
    (z-H_{0}^\rho)^{-1}=(z-X)^{-1}
    -\big(\rho+\ln (-z)\big)^{-1} (z-X)^{-1}|x^0\rangle\langle
x^0|(z-X)^{-1}.
  \end{equation*}
  In particular, $H_0^\infty=X$.

 Let $\rr\ni\tau\mapsto U_\tau$ be the  group of
dilations on $L^2[0,\infty[$, that is \[(U_\tau f)(x)=\e^{\tau/2} f(\e^\tau x).\]

We say that $B$ is   homogeneous of degree $\nu$ if
\[U_\tau B
U^{-1}_\tau= \e^{\nu\tau}B.\]

E.g., $X$ is homogeneous of degree $1$ and 
$|x^{\frac{m}{2}}\rangle\langle
x^{\frac{m}{2}}|$ is homogeneous of degree $1+m$.

The group of dilations (``the renormalization group'') acts on our
operators in a simple way:
\begin{align*}
  U_\tau H_{m,\lambda}U_\tau^{-1}&=\e^\tau H_{m,\e^{\tau
      m}\lambda},\\
  U_\tau H_{0}^\rho U_\tau^{-1}&=\e^\tau H_0^{\rho+\tau}.
\end{align*}

The essential spectrum of $H_{m,
  \lambda}$ and $H_0^\nu$ is $[0,\infty[$.
The point spectrum is more intricate, and is described by the following theorem:\pagebreak
\begin{theorem}\leavevmode \ben\item
$z\in\cc\backslash[0,\infty[$ belongs to the point spectrum of
    $H_{m,\lambda}$ iff 
    \[
    (-z)^{-m}=\lambda\frac{\pi}{\sin\pi m}.\]
\item
 $H_0^\rho$ possesses an eigenvalue iff $-\pi<\Im\rho<\pi$, and
  then it is $z=-\e^\rho$.
\een\end{theorem}

  For a given pair $(m,\lambda)$ all eigenvalues form a  geometric sequence that
lies
on   a  logarithmic spiral, which
should be viewed as a curve on the Riemann surface of the logarithm.
Only its ``physical sheet'' gives rise to eigenvalues.
For $m$
which are not purely imaginary, only a finite piece of the spiral is
on the ``physical sheet'' and therefore
the number of eigenvalues is finite.

If $m$ is purely imaginary, this spiral degenerates to a
 half-line starting at the origin.

 If $m$ is real, the spiral degenerates to a circle. But then the
 operator has at most
 one eigenvalue.
 
The following theorem about the number of eigenvalues of $H_{m,\lambda}$ is proven in~\cite{DR}. It describes an interesting pattern of ``phase transitions'' when we vary the parameter $m$. In this theorem, we denote by $\sp_\p(A)$ denotes the set of eigenvalues of an operator $A$ and $\#X$ denotes the number of elements of the set $X$.
\begin{theorem}
 Let $m= m_\r+\i m_\i \in \cc\backslash\{0\}$ with $|m_\r|<1$.
\begin{enumerate}
\item[(i)] Let $m_\r=0$.
  \begin{enumerate}\item[(a)] If $\frac{\ln(|\varsigma|)}{m_\i}\in ]-\pi,\pi[$,
then $\#\sp_\p(H_{m,\lambda}) = \infty$, \item[(a)] if
 $\frac{\ln(|\lambda\frac{\pi}{\sin\pi m}|)}{m_\i}\not \in ]-\pi,\pi[$
then $\#\sp_\p(H_{m,\lambda}) = 0$.\end{enumerate}
\item[(ii)]  If $m_\r\neq 0$ and if $N\in \nn$ satisfies
$N<\frac{m_\r^2+m_\i^2}{|m_\r|} \leq N+1$, then
\begin{equation*}
\#\sp_\p(H_{m,\lambda})\in \{N,N+1\}.
\end{equation*}
\end{enumerate}
\end{theorem}

\section{Homogeneous Schrödinger operators}
\label{sec2}

Let $\alpha\in\cc$. Consider the differential expression 
\begin{equation*}L_\alpha=-\partial_x^2+\Big(-\frac14+
\alpha\Big)\frac{1}{x^{2}}.\label{1}\end{equation*}
$L_\alpha$ is  is  homogeneous of
degree $-2$.
Following~\cite{BDG}, we would like to interpret $L_\alpha$ as a closed operator on $L^2[0,\infty[$
homogeneous of degree $-2$.

 $L_\alpha$, and closely related operators $H_m$ that we introduce
shortly,
 are interesting for many
reasons.
\begin{itemize}
\item They  appear
 as the  radial part of the Laplacian in all dimensions,
  in
  the decomposition of
  Aharonov-Bohm   Hamiltonian, in the membranes with  conical
   singularities, in the theory of many body systems with contact interactions and in the  Efimov effect.
\item They have rather subtle and rich properties illustrating various
  concepts of the  operator theory in Hilbert spaces (eg. the  Friedrichs and
 Krein
  extensions,  holomorphic families of closed operators).

\item Essentially all basic objects related to $H_m$, such as their
   resolvents, spectral projections, M{\o}ller and scattering operators, can be
  explicitly computed.
\item A number of
 nontrivial identities involving special functions, especially from the  Bessel
family,   find an
  appealing operator-theoretical interpretation in terms of the operators
  $H_m$. E.g. the Barnes identity
 leads to the formula for M{\o}ller
  operators. 
\end{itemize}

We start the Hilbert space theory of the operator  $L_\alpha$ by defining 
its two naive interpretations on $L^2[0,\infty[$:
\begin{enumerate} \item The  minimal operator $L_\alpha^{\min}$: We start from $L_\alpha$ on
$C_\c^\infty]0,\infty[$, and then we take its closure.
\item The maximal operator
$L_\alpha^{\max}$: We consider the domain consisting of all
  $f\in L^2[0,\infty[$ such that $L_\alpha f\in L^2[0,\infty[$.
\end{enumerate}

We will see  that it is often natural to write $\alpha=m^2$. Let us describe basic properties of $L_{m^2}^{\max}$ and $L_{m^2}^{\min}$:
\begin{theorem} \leavevmode\begin{enumerate}
\item For $1\leq\Re m$, $L_{m^2}^{\min}=L_{m^2}^{\max}$.
\item For $-1<\Re m<1$, 
 $L_{m^2}^{\min}\subsetneq L_{m^2}^{\max}$, and the codimension of
  their domains  is $2$.
\item $ (L_{\alpha}^{\min})^*=L_{\bar\alpha}^{\max}$.
Hence,  for $\alpha\in\rr$, $L_\alpha^{\min}$ is Hermitian.
\item $L_{\alpha}^{\min}$ and $L_{\alpha}^{\max}$ are homogeneous of
  degree $-2$.
\end{enumerate}
\end{theorem}

Let $\xi$ be a compactly supported cutoff equal $1$ around $0$.

Let $-1\leq \Re m $.
It is easy to check that $x^{\frac12+m}\xi$ belongs to $\Dom L_{m^2}^{\max}$. 
We
define the operator $H_m$ to be the restriction of $L_{m^2}^{\max}$ to
\[\Dom L_{m^2}^{\min}+\cc x^{\frac12+m}\xi.\]

The operators $H_m$ are in a sense more interesting than  $L_{m^2}^{\max}$ and $L_{m^2}^{\min}$:
\begin{theorem}\leavevmode \begin{enumerate}
\item For $1\leq\Re m$, $L_{m^2}^{\min}=H_m=L_{m^2}^{\max}$.
\item For $-1<\Re m<1$, 
 $L_{m^2}^{\min}\subsetneq H_m
\subsetneq L_{m^2}^{\max}$ and the  codimension of the  domains is $1$.
\item $ H_m^*=H_{\bar m}$. Hence,  for $m\in ]-1,\infty[$, $H_m$ is self-adjoint.
\item $H_m$ is homogeneous of
  degree $-2$.
\item $\sp H_m=[0,\infty[$.
\item $\{\Re m>-1\}\ni m\mapsto H_m$ is a holomorphic family of
  closed operators. \end{enumerate}
\end{theorem}

The theorem below is devoted to self-adjoint operators within the family~$H_m$.

\begin{theorem}\label{th:imain}
\leavevmode\begin{enumerate}
\item For $\alpha\geq 1$, $L_\alpha^{\min}=H_{\sqrt \alpha}$ is essentially
 self-adjoint
on $C_{\rm c}^\infty]0,\infty[$.
\item For $\alpha<1$,  $L_\alpha^{\min}$ is Hermitian but not essentially
  self-adjoint on $C_{\rm c}^\infty]0,\infty[$. It has deficiency indices $1,1$.
\item
 For $0\leq\alpha<1$,
 the operator $H_{\sqrt\alpha}$ is the  Friedrichs extension
 and
 $H_{-\sqrt\alpha}$ is the   Krein extension of 
$L_\alpha^{\min}$.
\item $H_{\frac12}$ is the Dirichlet Laplacian and
 $H_{-\frac12}$ is the  Neumann Laplacian on halfline.
\item
For $\alpha<0$,
$L_\alpha^{\min}$ has no homogeneous selfadjoint extensions.
\end{enumerate}
\end{theorem}

Various objects related to $H_m$
can be computed with help of  functions from the Bessel family.
Indeed, we have the following identity
\begin{equation*}
x^{-\frac12}
\Big(-\partial_x^2+\big(-\frac14+
\alpha\big)\frac{1}{x^{2}}\pm1\Big)x^{\frac12}
=-\partial_x^2-\frac1x\partial_x+\big(-\frac14+
\alpha\big)\frac{1}{x^{2}}\pm1,\end{equation*}
where the rhs defines the well-known (modified) Bessel equation.

One can compute explicitly the resolvent of $H_m$:
\begin{theorem}\label{th:reso}
Denote by $R_m(-k^2;x,y)$  the integral kernel of
the operator $(k^2+H_m)^{-1}$. Then for $\Re k>0$ we have
\[
R_m(-k^2; x,y) = 
\begin{cases} \sqrt{xy}I_m(kx)K_m(ky) & \text{ if }  x<y, \\
\sqrt{xy}I_m(ky)K_m(kx) & \text{ if }  x>y,  \end{cases}
\]
where $I_m$ is the  modified Bessel function and $K_m$ is the 
MacDonald function.
\end{theorem}

The operators $H_m$ are similar to self-adjoint operators. Therefore, they possess the spectral projection onto any Borel subset of their spectrum
$[0,\infty[$. In particular, below we give a formula for the spectral projection of $H_m$ onto the interval $[a,b]$:
\begin{proposition}\label{prop:hmproj} For $0<a<b<\infty$, the integral kernel of $\one_{[a,b]}(H_m)$ is
\begin{equation*}
\one_{[a,b]}(H_m)(x,y) 
 = \int_{\sqrt a}^{\sqrt b} \sqrt{xy}J_m(k x) J_m(k y)k\d k,\label{hankel2}
\end{equation*}
where $J_m$ is the Bessel function.
\end{proposition}

One can diagonalize the operators $H_m$ in a natural way, using the so-called Hankel transformation $\cF_m$, which is
 the operator on $L^2[0,\infty[$ given by
\begin{equation}
\big(\cF_mf\big)(x):= \int_0^\infty J_m(kx)\sqrt{kx}f(x)\d x
\label{hankel}\end{equation}

\begin{theorem} \label{th:hank} $\cF_m$ is a bounded invertible involution on
$L^2[0,\infty[$ diagonalizing $H_m$, more precisely
$$
\cF_mH_m\cF_m^{-1}=X^2.
$$
It satisfies $\cF_mA=-A\cF_m$,
where
\[A=\frac1{2\i} (x\partial_x+\partial_xx)\]
is the self-adjoint  generator of dilations.
\end{theorem}

It turns out that the Hankel transformation can be expressed in terms of the generator of dilations. This expression, together with the Stirling formula for the asymptotics of the Gamma function, proves
the boundedness of $\cF_m$.

\begin{theorem}
  Set 
   \[\cI f(x)=x^{-1}f(x^{-1}),\quad
\Xi_m(t)=\e^{\i\ln(2)t}\frac{\Gamma(\frac{m+1+\i t}{2})}{\Gamma(\frac{m+1-\i t}{2})}.
\]Then
\[\cF_m=\Xi_m(A)\cI.\]
Therefore,
 we have the identity
 \begin{equation}
   H_m:=\Xi_m^{-1}(A) X^{-2}\Xi_m(A).\label{ham}
 \end{equation}
\end{theorem}

(Result  obtained independently by Bruneau, Georgescu,
and myself in~\cite{BDG}, and 
 by Richard and Pankrashkin in~\cite{PR}).

The operators $H_m$ generate 1-parameter groups of bounded operators. They possess scattering theory and one can explicitly compute their M{\o}ller (wave) operators and the scattering operator.

\begin{theorem}
The M{\o}ller operators associated to the
pair $H_m,H_k$ exist and
\begin{equation*}
  \Omega_{m,k}^\pm \,:=\,\lim_{t\to\pm\infty}\e^{\i tH_m}\e^{-\i tH_k}
=\e^{\pm \i(m-k)\pi/2}\cF_m\cF_k
=
\e^{\pm \i(m-k)\pi/2}\frac{\Xi_k(A)}{\Xi_m(A)}.
\label{pankra}\end{equation*}
\end{theorem}

The formula
(\ref{ham}) valid for 
$\Re m>-1$
can be used as an alternative definition of the family $H_m$ also beyond this domain.
It defines a family of closed operators for the
parameter $m$ that belongs to
 \beq \{m\mid \Re m\neq-1,-2,\dots\}\cup\rr.\label{doma}\eeq
 Their spectrum is always equal to $[0,\infty[$
    and they are analytic in the interior of (\ref{doma}).

In fact, $\Xi_m(A)$ is a unitary
    operator for all real values of $m$.
Therefore, for $m\in\rr$, (\ref{ham}) is well-defined and self-adjoint.

$\Xi_m(A)$ is
bounded and invertible also for all $m$ such that $\Re
m\neq-1,-2,\dots$. Therefore, formula (\ref{ham}) defines an
operator for all such $m$.

One can then pose various questions: 
\begin{itemize}
\item
What happens with these operators along the lines $\Re m=-1,-2,\dots$?
\item What is the meaning of these operators to the left of $\Re=-1?$ (They are not differential operators!)
\end{itemize}

Let us describe a certain precise conjecture about the family $H_m$. In order to state it we need to define the concept of a holomorphic family of closed operators.

The definition (or actually a number of equivalent definitions) of a
 holomorphic family of bounded operators is quite obvious and
does not need to be recalled. In the case of unbounded operators the
situation is more subtle, and is described e.g., in~\cite{Kato2}, see also~\cite{DW}.

Suppose that $\Theta$ is an open subset of $\cc$, $\cH$ is a Banach
space, and $\Theta\ni z\mapsto H(z)$ is a function  whose values
are closed operators on $\cH$.  We say that this is a 
  holomorphic family of closed operators if for each $z_0\in\Theta$
there exists a neighborhood $\Theta_0$ of $z_0$, a Banach space
$\cK$ and a holomorphic family of injective bounded operators $\Theta_0\ni
z\mapsto B(z)\in B(\cK,\cH)$ such that $\Ran B(z)=\cD(H(z))$ and
\begin{equation*}
\Theta_0\ni z\mapsto H(z)B(z)\in B(\cK,\cH)\label{holo2}
\end{equation*}
is a holomorphic family of bounded operators.

We have the following practical criterion:

\begin{theorem}\label{crit} Suppose that $\{H(z)\}_{z\in\Theta}$ is a
function whose values are closed operators on $\cH$. Suppose in
addition that for any $z\in\Theta$ the  resolvent set of $H(z)$
is  nonempty. Then $z\mapsto H(z)$ is a  holomorphic family of
closed operators if and only if for any $z_0\in\Theta$ there exists
$\lambda\in\C$ and a neighborhood $\Theta_0$ of $z_0$ such that
$\lambda$ belongs to the resolvent set of $H(z)$
 for $z\in\Theta_0$ and $z\mapsto
(H(z)-\lambda)^{-1}\in B(\cH)$ is holomorphic on $\Theta_0$.
\end{theorem}

The above theorem indicates that it is more difficult to study
holomorphic families of closed operators that for some values of the
complex parameter have an  empty resolvent set.
We have the following conjecture (formulated as an open question in~\cite{BDG}), so far unproven:

\begin{conjecture}
It is impossible to extend \[\{\Re m>-1\}\ni m\mapsto H_m\]
to 
a   holomorphic family of closed operators on a larger connected open subset of
$\cc$.
\end{conjecture}

\section{Almost homogeneous Schrödinger operators}
\label{sec3}

For $-1<\Re m<1$ the codimension of $\Dom(L_{m^2}^{\min})$ in $\Dom(L_{m^2}^{\max})$
is two. Therefore, following~\cite{DR}, one can fit a 1-parameter family of closed operators in between $L_{m^2}^{\min}$ in $L_{m^2}^{\max}$, mixing the boundary condition
$x^{\frac12+m}$ and $x^{\frac12-m}$. These operators in general are no longer homogeneous---their homogeneity is broken by the boundary condition. We will say that they are {\em almost homogeneous}.

More precisely, 
for any $\kappa\in \C\cup \{\infty\}$ let  $H_{m,\kappa}$
be the restriction of $L_{m^2}^{\max}$ to the domain
\begin{multline*}
\Dom(H_{m,\kappa})  = \big\{f\in \Dom(L_{m^2}^{\max})\mid
 \hbox{ for some }  c \in \C,\\
f(x)- c
\big(x^{1/2-m} +\kappa
x^{1/2+m}\big)\in\Dom(L_{m^2}^{\min})
\text{ around }
0\big\},\qquad\kappa\neq\infty;
\end{multline*}
\begin{multline*}
\Dom(H_{m,\infty})  =\big\{f\in \Dom(L_{m^2}^{\max})\mid
 \hbox{ for some }  c \in \C,\\
f(x)- c
x^{1/2+m}\in\Dom(L_{m^2}^{\min})\text{ around } 0\big\}.
\end{multline*}

The case $m=0$ needs a special treatment.
For
$\nu\in\C\cup \{\infty\}$, let $H_0^\nu$ be the restriction of
$L_0^{\max}$ to
\begin{multline*}
\Dom(H_0^\nu)  = \big\{f\in \Dom(L_{0}^{\max})\mid
 \hbox{ for some }  c \in \C,\\
f(x)- c
\big(x^{1/2}\ln x +\nu x^{1/2}\big)\in\Dom(L_0^{\min})
\text{ around } 0\big\},\qquad\nu\neq\infty; 
\end{multline*}
\begin{multline*}
\Dom(H_0^\infty)  = \big\{f\in \Dom(L_0^{\max})\mid
 \hbox{ for some }  c \in \C,\\
f(x)- c
x^{1/2}\in\Dom(L_0^{\min})\hbox{ around }
0\big\}.
\end{multline*}

Here are the basic properties of almost homogeneous Schrödinger operators.
\begin{proposition}\leavevmode
\begin{enumerate}
\item For any $|\Re(m)|<1$, $\kappa\in \C\cup \{\infty\}$ 
\begin{equation*}\label{Eq_duality}
H_{m,\kappa}=H_{-m,\kappa^{-1}}.
\end{equation*}
\item
$H_{0,\kappa}$ does not depend on $\kappa$,
and these operators coincide with $H_0^\infty$.
\item We have
\begin{align*}
U_\tau H_{m,\kappa}U_{-\tau}&=\e^{-2\tau}H_{m,\e^{-2\tau m}\kappa},\\
U_\tau H_0^\nu U_{-\tau}&=\e^{-2\tau}H_0^{\nu+\tau}.
\end{align*}
In particular, only
\begin{equation*}
H_{m,0}=H_{-m},\quad
H_{m,\infty}=H_m,\quad
H_0^\infty=H_0
\end{equation*} are homogeneous.
\end{enumerate}\label{basic}
\end{proposition}

The following proposition describes self-adjoint cases among these operators.
\begin{proposition}.
\begin{equation*}\label{Eq_adjoint}
H_{m,\kappa}^*=H_{\bar m,\bar\kappa}\qquad \hbox{ and }\qquad
H_0^{\nu*} = H_0^{\bar \nu}.
\end{equation*}
In particular,
\begin{enumerate}
\item[(i)]  $H_{m,\kappa}$ is self-adjoint for $m\in]-1,1[$ and $\kappa\in\R\cup\{\infty\}$,
     and for $m\in \i\R$ and $|\kappa|=1$.
\item[(ii)]  $H_0^\nu$
is self-adjoint for
$\nu\in \R\cup \{\infty\}$.
\end{enumerate}\end{proposition}
The essential spectrum of  $H_{m,\kappa}$ and $H_0^\nu$ is always $[0,\infty[$. The following proposition describes the point spectrum in the self-adjoint case.
\begin{proposition}\leavevmode
  \begin{enumerate}\item         If $m\in]-1,1[$ and $\kappa
       \geq0$ or $\kappa=\infty$, then $H_{m,\kappa}$  has no eigenvalues.\item
       If $m\in]-1,1[$ and $\kappa<0$, then $H_{m,\kappa}$  has a single eigenvalue\\ at
       $-4\big(\frac{\Gamma(m)}{\kappa\Gamma(-m)}\big)^{\frac1m}$.
   \item If $m\in\i\R$ and
$|\kappa|=1$, then $H_{m,\kappa}$  has an infinite sequence of eigenvalues accumulating at $-\infty$ and $0$. If $m=\i m_\I$ and
    $ \e^{\i\alpha}=\frac{\kappa\Gamma(- \i m_\I)}{\Gamma(\i m_\I)}$, then
     these eigenvalues are $-4\exp(-\frac{\alpha+2\pi n}{m_\I})$, $n\in\zz$. \end{enumerate}\label{table0}
\end{proposition}

It is interesting to analyze how the set of self-adjoint extensions of
the Hermitian operator 
\begin{equation*}L_\alpha^{\min}=-\partial_x^2+\Big(-\frac14+
\alpha\Big)\frac{1}{x^{2}}\end{equation*}
depends on the real parameter $\alpha$.
Self-adjoint extensions form  a set isomorphic either to a point or to a circle. The ``renormalization group'' acts on this set by a continuous flow, as described by Proposition \ref{basic}.
This flow may have  fixed points.

The following table describes the various ``phases'' of the theory of self-adjoint extensions of $L_\alpha^{\min}$. To each phase I give a name inspired by condensed matter physics. The reader does not have to take these names very seriously, however I suspect that they have some deeper meaning.
\bigskip
\\
\begin{tabular}{llll}\hline\\
  $1\leq\alpha$&``gas''&point&\parbox{0.42\linewidth}{Unique fixed point: Friedrichs extension=Krein extension.}\\\\
  \hline\\
  $0<\alpha<1$&``liquid''&circle&\parbox{0.42\linewidth}
  {Two fixed points: Friedrichs and Krein extension.\\
    Ren. group flows from Krein to Friedrichs.\\
    On one semicircle of non-fixed points all have one bound state; on the other all have no bound states.}\\\\\hline\\
  $\alpha=0$&\parbox{0.25\linewidth}{``liquid--solid\\ phase transition''}&circle&\parbox{0.42\linewidth}
  {Unique fixed point: Friedrichs extension=Krein extension.\\ Ren. group flows from Krein to Friedrichs.\\
  Non-fixed points have one bound state.}\\\\
  \hline\\
    $\alpha<0$&``solid''&circle&\parbox{0.42\linewidth}
    {No fixed points.\\Ren. group rotates the circle.\\ All have infinitely many bound states.}\\\\
  \hline\\
\end{tabular}

\noindent The above table can be represented by the following picture, hopefully self-explanatory:

\medskip

\noindent
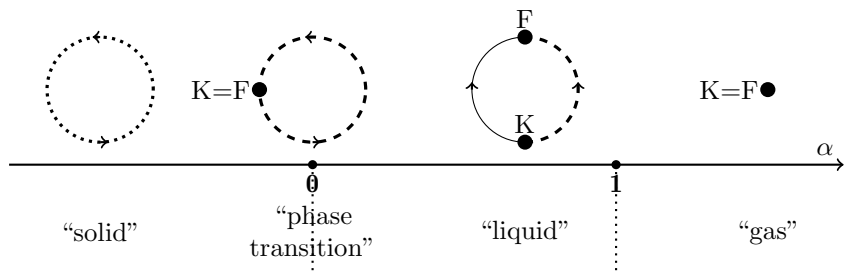
\begin{tikzpicture}
  
  \draw[very thick,dotted](0.2,1)circle(0.7);
  \draw[thick,->](0.2,1.7)--(0.1,1.7) ;
  \draw[thick,->](0.2,0.3)--(0.3,0.3) ;

  \draw[very thick,dashed](3,1)circle(0.7);
\fill[black](2.3,1)circle(0.1) node[anchor=east] {K=F};
  \draw[thick,->](3,1.7)--(2.9,1.7) ;
  \draw[thick,->](3,0.3)--(3.1,0.3) ;

  \draw[very thick,dashed](5.8,0.3)arc(-90:90:0.7);
    \draw[thin](5.8,1.7)arc(90:270:0.7);
  \fill[black](5.8,1.7)circle(0.1) node[anchor=south] {F};
  \fill[black](5.8,0.3)circle(0.1) node[anchor=south] {K};
  \draw[thick,->](5.1,1)--(5.1,1.1) ;
    \draw[thick,->](6.5,1)--(6.5,1.1) ;

    \fill[black](9,1)circle(0.1) node[anchor=east] {K=F};
    
  \draw[thick,-] (-1,0) -- (3,0) node[anchor=north] {\bf 0};
\fill[black] (3,0) circle (0.4ex);
\draw[thick,-] (3,0) -- (7,0) node[anchor=north] {\bf 1};
\fill[black] (7,0) circle (0.4ex);
\draw[thick,->] (7,0) -- (10,0)
 node[anchor=south east] {$\alpha$};

 \node at (0.2,-0.9){``solid''};
 \node at (3,-0.7){``phase};
 \node at (3,-1.1){transition''};
  \node at (5.8,-0.9){``liquid''};
  \node at (9,-0.9){``gas''};

  \draw[thick,dotted,-](3,0) --(3,-1.5);
    \draw[thick,dotted,-](7,0) --(7,-1.5);
\end{tikzpicture}

\bigskip

There exists a close link between almost homogeneous Schrödinger  operators described in this section and the ``toy model of renormalization group'' described in Section~\ref{sec1}. It turns out that the corresponding operators are similar to one another.

    Define the unitary operator
    \[ (If)(x):=x^{-\frac14}f(2\sqrt x).\]
    Its inverse is
        \[
        (I^{-1}f)(x):=\Big(\frac{y}{2}\Big)^{\frac12}f\Big(\frac{y^2}{4}\Big).\]
        Note that
        \begin{equation*}
          I^{-1}XI=\frac{X^2}{4},\quad
          I^{-1}AI=\frac{A}{2}.
        \end{equation*}

        We change slightly notation:
        the operators $H_m$, $H_{m,\kappa}$ and $H_0^\nu$ of this section
        will be denoted
        $\tilde H_m$, $\tilde H_{m,\kappa}$ and $
        \tilde H_0^\nu$.
Recall  that in (\ref{hankel}) we
        introduced the Hankel transformation $\cF_m$, which is a
        bounded invertible involution satisfying
        \begin{eqnarray*}\cF_m\tilde H_m\cF_m^{-1}&=&X^2,\\
          \cF_mA\cF_m^{-1}&=&-A.
          \end{eqnarray*}

Recall also that in Section~\ref{sec1} we introduced the operators $H_{m,\lambda}$ and $H_0^\rho$.

        The following theorem is proven in~\cite{D}:
        \begin{theorem}.
          \ben\item
If        $
\lambda\frac{\pi}{\sin(\pi m)}=\kappa\frac{\Gamma(m)}{\Gamma(-m)  },$
        then the operators $H_{m,\lambda}$ are similar to $\tilde H_{m,\kappa}$:
        \[
        \cF_m^{-1}I^{-1} H_{m,\lambda}I\cF_m=\frac14\tilde
        H_{m,\kappa},\]
            \item
If  $\rho=-2\nu$, then the operators $H_0^\rho$ are similar to $\tilde H_0^\nu$:
              \[
        \cF_m^{-1}I^{-1} H_0^\rho I\cF_m=\frac14\tilde
        H_0^\nu,\]
\een
\end{theorem}

\end{document}